\pdfoutput=1
\documentclass[amssymb,amsmath,prl,twocolumn,superscriptaddress]{revtex4-1}

\usepackage[caption=false]{subfig}
\usepackage{graphicx}
\usepackage{bm}
\usepackage{amssymb,amsmath}
\usepackage{mathrsfs}
\usepackage{latexsym}
\usepackage{color}
\usepackage[normalem]{ulem}
\usepackage{dcolumn}
\usepackage[colorlinks,urlcolor=dodgerblue,citecolor=dodgerblue,linkcolor=dodgerblue,pdfusetitle]{hyperref}
\usepackage[usenames,dvipsnames]{xcolor}
\usepackage[utf8]{inputenc}
\usepackage{microtype}
\usepackage{etoolbox}
\usepackage{accents}
\usepackage{multirow}

\allowdisplaybreaks

\definecolor{ZurichBlue}{rgb}{.255,.41,.884} 		
\definecolor{ZurichRed}{rgb}{0.9, 0.1, 0} 			
\definecolor{ZurichGreen}{rgb}{.196,.504,.396} 		
\definecolor{ZurichYellow}{rgb}{1,.648,0} 			
\definecolor{dodgerblue}{rgb}{0.12, 0.56, 1.0}
\definecolor{azure}{rgb}{0.0, 0.5, 1.0}
\definecolor{alizarincrimson}{rgb}{0.82, 0.1, 0.26}
\definecolor{mediumpurple}{rgb}{0.58, 0.44, 0.86}
\definecolor{lasallegreen}{rgb}{0.03, 0.47, 0.19}
\definecolor{my_gray}{rgb}{0,0,0}
\definecolor{uniwienblue}{HTML}{006699}
\definecolor{walesred}{HTML}{ff0038}
\definecolor{myorange}{HTML}{FF6C0C}
\definecolor{ferngreen}{HTML}{009246}
\definecolor{scarletred}{HTML}{CD212A}

\definecolor{dodgerblue}{HTML}{1E90FF}
\definecolor{viennared}{HTML}{DA0A14}
\definecolor{ctorange}{HTML}{FF6C0C}
\definecolor{wales}{HTML}{ff0038}
\definecolor{benettongreen}{HTML}{009421}
\definecolor{ferrarired}{HTML}{ff2800}
\definecolor{austriawienpurple}{HTML}{441678}
\definecolor{gray}{HTML}{F0F0F0}
\definecolor{LightCyan}{rgb}{0.88,1,1}
\newcolumntype{a}{>{\columncolor{gray}}c}
\newcolumntype{b}{>{\columncolor{white}}c}

\hypersetup{
     colorlinks=true,
     linkcolor=dodgerblue,
     filecolor=dodgerblue,
     citecolor = dodgerblue,      
     urlcolor=dodgerblue,
     }

\newcommand{\texi}{\theta_{{\rm{ex}},i}}
\newcommand{\tini}{\theta_{{\rm{in}},i}}

\newcommand{\beq}{\begin{equation}}
\newcommand{\eeq}{\end{equation}}

\makeatletter
\def\maketitle{\@author@finish \title@column\titleblock@produce \suppressfloats[t]}
\makeatother

\begin{document}

\title{Impact of Dynamical Tides on the Reconstruction of the Neutron Star Equation of State}

\newcommand{\BHam}{\affiliation{School of Physics and Astronomy and Institute for Gravitational Wave Astronomy, University of Birmingham, Edgbaston, Birmingham, B15 2TT, United Kingdom}}

\author{Geraint Pratten}
\email{G.Pratten@bham.ac.uk}
\BHam

\author{Patricia Schmidt}
\email{P.Schmidt@bham.ac.uk}
\BHam

\author{Natalie Williams}
\email{nxw049@bham.ac.uk}
\BHam

\hypersetup{pdfauthor={Pratten, Schmidt}}


\begin{abstract}
Gravitational waves (GWs) from inspiralling neutron stars afford us a unique opportunity to infer the as-of-yet unknown equation of state of cold hadronic matter at supranuclear densities. During the inspiral, the dominant matter effects are arise due to the star's response to their companion's tidal field, leaving a characteristic imprint in the emitted GW signal. This unique signature allows us to constrain the cold neutron star equation of state. At GW frequencies above $\gtrsim 800$Hz, however, subdominant tidal effects known as dynamical tides become important. In this letter, we demonstrate that neglecting dynamical tidal effects associated with the fundamental ($f$-) mode leads to large systematic biases in the measured tidal deformability of the stars and hence in the inferred neutron star equation of state. Importantly, we find that $f$-mode dynamical tides will already be relevant for Advanced LIGO's and Virgo's fifth observing run ($\sim 2025$) -- neglecting dynamical tides can lead to errors on the neutron radius of $\mathcal{O}(1{\rm km})$, with dramatic implications for the measurement of the equation of state. 
Our results demonstrate that the accurate modelling of subdominant tidal effects beyond the adiabatic limit will be crucial to perform accurate measurements of the neutron star equation of state in upcoming GW observations.
\end{abstract}

\pacs{%
  04.30.-w,  
  04.80.Nn, 
  04.25.D-,  
  04.25.dg   
  04.25.Nx,  
}

\maketitle

\section{Introduction}
\label{sec:intro}
The observations of gravitational waves (GWs) from the inspiral of binary neutron stars (BNS), GW170817~\cite{GW170817} and GW190425~\cite{Abbott:2020uma}, has opened up a new way of studying ultradense nuclear matter. Gravitational waves carry characteristic information about the composition of neutron stars, providing a unique means to infer their elusive equation of state (EOS)~\cite{Flanagan:2007ix}.
Observations with the current ground-based GW detector network of Advanced LIGO~\cite{TheLIGOScientific:2014jea} and Advanced Virgo~\cite{TheVirgo:2014hva} have yielded the first direct constraints on the EOS~\cite{GW170817EOS,LIGOScientific:2019eut}, with recent complementary information obtained from pulsar observations~\cite{Bogdanov:2021yip} and terrestrial experiments~\cite{Reed:2021nqk}. 

A precision measurement of the nuclear EOS from inspiralling BNS is a prime science objective for the next-generation of ground-based (A+) \cite{Miller:2014kma,Aasi:2013wya} and third-generation (3G) GW detectors such as the Einstein Telescope~\cite{Maggiore:2019uih} and Cosmic Explorer~\cite{Reitze:2019iox}. In order the make such a measurement from GW observations, accurate theoretical models of the emitted GW signal are required. GWs from inspiralling neutron stars (NS) differ from those of binary black holes (BBHs) due to finite-size (tidal) effects, which manifest themselves as an additional phasing term~\cite{Flanagan:1997fn}. This imprint arises from an additional, tidally induced quadrupole moment sourced by their companion's gravitational field, which enhances GW emission. 
The dominant (Newtonian) tidal contributions to the GW phase are quadrupolar ($\ell =2$) tidal effects arising from the excitation of the star's fundamental oscillation mode ($f$-mode), which depend indirectly on the EOS through the macroscopic (dimensionless) tidal deformability parameter $\Lambda^A_\ell$ and the (dimensionless) $f$-mode angular frequency $\Omega^A_\ell$ of the $A$-th neutron star for the $\ell$-th multipolar mode. In the adiabatic limit, i.e. where the $f$-mode frequency is much larger than the orbital frequency, tidal effects in the GW phase are governed by the (dimensionless) binary tidal deformabilities $\tilde{\Lambda}$ and $\delta \tilde{\Lambda}$~\cite{Flanagan:2006sb, Favata:2013rwa, Wade:2014vqa}, independent of the $f$-mode frequency. During the late inspiral, however, dynamical tidal effects induced by the resonant excitation of the $f$-mode, which explicitly depend on the $f$-mode frequency, must be taken into account in order to accurately describe the GW signal~\cite{Hinderer:2016eia, Schmidt:2019wrl}. 

To date, much emphasis has been placed on the accurate modelling of the point-particle phase~\cite{Messina:2019uby,Samajdar:2019ulq} and on adiabatic tidal effects~\cite{Akcay:2018yyh,Nagar:2018plt,Henry:2020ski,Gamba:2020wgg, Narikawa:2021pak}. In recent years, however, much progress has been made in including dynamical and non-perturbative tidal effects into waveform models~\cite{Hinderer:2016eia, Steinhoff:2016rfi, Schmidt:2019wrl,Ma:2020rak,Poisson:2020eki,Dietrich:2019kaq,Pratten:2019sed,Andersson:2019dwg}. 
In this letter we demonstrate for the first time that dynamical tides, despite being a higher-order tidal effect, play a key role for a precision measurement of the NS EOS. 
We show that neglecting dynamical tides leads to significant systematic biases in the recovered tidal deformability $\tilde{\Lambda}$ and consequently in the inference of the neutron star EOS from individual BNS observations as well as entire populations. 
Crucially, we show that such systematic biases are already problematic for the A+ detector network scheduled to commence observing in $\sim 2025$: For a semi-realistic population of BNS in the A+ network, we estimate that systematic errors on the inferred radius of NS could be as large as $\sim \mathcal{O}(1{\rm{km}})$. Our results highlight the urgent need for more sophisticated waveform models that accurately model higher-order tidal effects and are computationally efficient, as well as a more robust treatment of waveform systematics in order to accurately measure the neutron star EOS in upcoming GW observations.

\section{Methods}
\label{sec:methods}
We consider two different network configurations: First a network consisting of the two LIGO detectors and Virgo, operating at the A+ design and the low-limit sensitivity~\cite{O5PSDs} respectively, as anticipated for the fifth observing run (O5)~\cite{Aasi:2013wya}. Secondly, we consider a single triangular ET detector with the proposed ET-D sensitivity \cite{Hild:2010id}.

We model the simulated BNS signals using the inspiral-only frequency-domain \texttt{TaylorF2} waveform approximant with a point-particle phase up to 3.5 post-Newtonian (PN) order \cite{Pratten:2020fqn}, adiabatic tidal effects up to 7.5PN \cite{Flanagan:2007ix,Vines:2011ud,Damour:2012yf} and quadrupolar ($\ell=2$) dynamical tidal effects at 8PN \cite{Schmidt:2019wrl}, as implemented in the LIGO Algorithms Library~\cite{LAL}. We assume that the neutron stars are nonspinning and undergo a quasi-circular inspiral for consistency with the dynamical tides phase model but note that spins and eccentricity will further enhance the excitation of the $f$-modes~\cite{Doneva:2013zqa, Steinhoff:2021dsn, Chirenti:2016xys}. The BNS waveforms start at a GW frequency of $f=20$ Hz and are truncated at either the frequency of the innermost stable circular orbit or the contact frequency~\cite{Dietrich:2018uni}.

To determine the impact of dynamical tides on the measurement of the EOS, we perform Bayesian inference using the nested sampling~\cite{skilling2006nested} algorithm \texttt{dynesty}~\cite{Speagle:2019ivv} as implemented in the GW inference package \texttt{Bilby}~\cite{Ashton:2018jfp}. We recover the binary parameters with and without the inclusion of dynamical tides, using quasi-universal relations~\cite{Chan:2014kua} to determine the $f$-mode frequencies. Whilst we do not consider any errors in the universal relations, we note that a more detailed and careful understanding will be required for precision measurements when using such relations \cite{Godzieba:2020bbz}. We further assume that the binaries can be associated to an EM counterpart allowing us to fix the luminosity distance, right ascension and declination when performing parameter estimation; we marginalise over coalescence time and phase and do not include Gaussian noise or calibration uncertainties in our analyses. We adopt priors that are uniform in the component masses $m_i$ but sample in chirp mass $\mathcal{M}_c$ and mass ratio $q = m_1/m_2$ \cite{Veitch:2014wba}, and uniform priors on the binary tidal deformability parameters $\tilde{\Lambda}$ and $\delta{\tilde{\Lambda}}$~\cite{Wade:2014vqa}. 

Assuming that all NS have a universal EOS, we study the impact of dynamical tides on inferences made from a semi-realistic population of $N$ BNS.
Here we consider two complementary approaches to inferring the EOS using the relation between the mass and tidal deformability, $\Lambda (m)$, and the relation between the mass and radius, $R(m)$, respectively. In the first approach, we estimate the impact of dynamical tides on measuring $\Lambda(m)$. Following \cite{DelPozzo:2013ala}, we perform a Taylor expansion of $\Lambda (m)$ in $(m - m_c)/M_{\odot}$ to linear order. As highlighted in \cite{DelPozzo:2013ala}, higher order terms are poorly constrained and we can only accurately measure the coefficients for a fiducial mass. In addition, it is difficult to impose meaningful \textit{a priori} information on the functional form of $\Lambda(m)$ \cite{Lattimer:2014sga,Lackey:2014fwa}. Adopting a canonical neutron star mass of $m_c=1.33\, M_\odot$, we determine $\Lambda_{1.33}$ using the mass and tidal posterior samples from each event in the population
\begin{equation}
\label{eq:stacked}
    \Lambda_{1.33} \simeq (1.33 - m_2)\frac{\Lambda_2 - \Lambda_1}{m_2 - m_1} + \Lambda_2.
\end{equation}
The joint likelihood for $N$ binaries is given by
\begin{equation}
    \mathcal{L}(\Lambda_{1.33}) \sim \prod_{n=1}^{N} p(\Lambda_{1.33}|d_n, H_i) \, p(\Lambda_{1.33}|H_i)^{-1},
\end{equation}
where $H_i$ denotes either the adiabatic or dynamical tides hypothesis and $d_i$ the segment of data for the $i$-th binary. 

In the second approach, we perform Bayesian inference to directly reconstruct the EOS and gauge the impact of neglecting dynamical tides on the inference of macroscopic NS properties. Following \cite{Read:2008iy,Lackey:2014fwa}, we adopt a piecewise polytropic model $p(\rho) = K_i \rho^{\Gamma_i}$, where $\Gamma_i$ are the adiabatic indices and $K_i$ enforces that the pressure $p$ is continuous at the boundaries between density intervals $\rho_{i-1} < \rho < \rho_i$. We adopt the density intervals reported in \cite{Lackey:2014fwa}, which are chosen to minimize the least-squares error between the piecewise polytropic fits and tabulated theoretical EOSs \cite{Read:2008iy}. We use three adiabatic indices $\lbrace \Gamma_1, \Gamma_2, \Gamma_3 \rbrace$ and a constant $p_1 = \rho(p_1)$ that sets the overall pressure scale. An advantage to this framework is that we can impose \textit{a priori} constraints on the EOS that are otherwise difficult to incorporate into the $\Lambda (m)$ fit. Explicitly, we require that the EOS is i) thermodynamically stable and hence a monotonically increasing function $d p / d \epsilon \geq 0$, where $\epsilon$ is the energy density, ii) must obey causality constraints, $v_s = \sqrt{dp/d\epsilon} < c$, and iii) that the maximum supported NS mass is compatible with the heaviest known pulsar, PSR J0740+6620, $M_{\rm max} \geq 2.08 M_{\odot}$ \cite{Fonseca:2021wxt}. We infer the EOS parameters from the population of $N$ binaries following the framework outlined in \cite{Lackey:2014fwa}. Recycling the posterior samples from earlier, we can construct a pseudo-likelihood for the intrinsic parameters of each event $\theta_{{\rm{in}}} = \lbrace \mathcal{M}, q, \mathcal{E} \rbrace$ by marginalizing over all extrinsic and nuisance parameters $\theta_{{\rm{ex}}}$ \cite{Lackey:2014fwa}
\begin{align}
    \mathcal{L}(d_i, \tini, \mathcal{H}, \mathcal{I}) = \displaystyle\int d \texi \, p(\texi | \mathcal{H}, \mathcal{I} ) \, p(d_i | \theta_i, \mathcal{H}, \mathcal{I}) .
\end{align}
\newline
where $\mathcal{H}$ denotes the waveform model used and $\mathcal{I}$ denotes all background information for the EOS and waveform parameters. This allows us to construct a marginalized posterior density function for the EOS parameters $\mathcal{E} = \lbrace p_1, \Gamma_1, \Gamma_2, \Gamma_3 \rbrace$ by constructing the joint-likelihood for all $N$ BNS events, re-expressing $\tilde{\Lambda}_i$ in terms of the EOS parameters and $\lbrace \mathcal{M}, q, \mathcal{E} \rbrace$, and marginalizing over the masses using MCMC algorithms \cite{Lackey:2014fwa}
\begin{align}
\label{eq:eos_direct}
    &p(\mathcal{E} | d_N, \mathcal{H}, \mathcal{I}) = \frac{1}{p(d_N | \mathcal{H}, \mathcal{I})} \int d \mathcal{M} \, d q \, p(\mathcal{E} | \mathcal{H}, \mathcal{I}), \\
    \nonumber &\; \times \displaystyle\prod_{i=1}^N  p(\mathcal{M}_i,q_i | \mathcal{E}, \mathcal{H}, \mathcal{I}) \mathcal{L}(d_i, \tini, \mathcal{H}, \mathcal{I})\Big|_{\tilde{\Lambda}_i = \tilde{\Lambda}(\mathcal{M}_i,\eta_i,\textrm{EOS})} ,
\end{align}

where $d \mathcal{M} = d \mathcal{M}_1 \cdots d \mathcal{M}_N$, and $d_q = d {q_1} \cdots d q_N$. The term $p(\mathcal{M}_i,q_i | \mathcal{E}, \mathcal{H}, \mathcal{I} )$ is a conditional prior that captures the range of masses supported by an EOS with parameters  $\mathcal{E}$. Finally, we construct posterior distributions for the EOS parameters by marginalizing over the mass parameters using nested sampling~\cite{Speagle:2019ivv}. Using the posteriors for the EOS parameters $\mathcal{E}$, we reconstruct the posterior distributions for the macroscopic relation $R(m)$ as well as $\Lambda (m)$.  

\section{Results}
\label{sec:}
\subsection{Systematic series}
To assess the impact of systematic biases in measurements of the tidal deformability induced by neglecting dynamical tidal effects, we first consider a series of simulated BNS with fixed mass ratio $q=0.855$ and varying total mass $M=m_1 + m_2$ and hence varying chirp mass $\mathcal{M}_c = (m_1 m_2)^{3/5}/M^{1/5}$ and tidal binary deformability $\tilde{\Lambda}$. The mass and mass ratio are chosen to be broadly consistent with both GW170817~\cite{GW170817} and GW190425~\cite{Abbott:2020uma}.
Tidal effects are related to the compactness of the neutron star, with lighter neutron stars having larger tidal deformabilities for a fixed EOS. The quadrupolar $f$-mode contribution to the GW phasing scales as $\delta \varphi_{\rm{dyn}} \propto (\Lambda_{2A}) / (\Omega_{2A}^2)$, where lighter NS have lower $f$-mode frequencies and hence a stronger excitation of the $f$-modes. Tidal effects, both adiabatic and dynamical, become suppressed as we simultaneously increase $\mathcal{M}_c$ and decrease $\tilde{\Lambda}$.

We adopt a soft EOS (APR4~\cite{Akmal:1998cf}), consistent with current observations \cite{LIGOScientific:2019eut}. Further, for this series, we distribute the binaries at a distance between 95 and 143 Mpc such that the signal-to-noise ratio (SNR) in the triangular ET-D configuration is fixed to $\rho_{\rm{ET}} \simeq 500$, where we expect dynamical tides to be distinguishable even for heavy NS~\cite{Williams:2022vct}. For the LIGO-Virgo A+ network, this translates into a fixed network SNR of $\rho_{\rm A+} \simeq 50$. We also adopt a fixed small inclination angle of $\iota = 12.6^\circ$, consistent with GW170817, and fix the sky location. 

\begin{figure}[t!]
\includegraphics[width=\columnwidth]{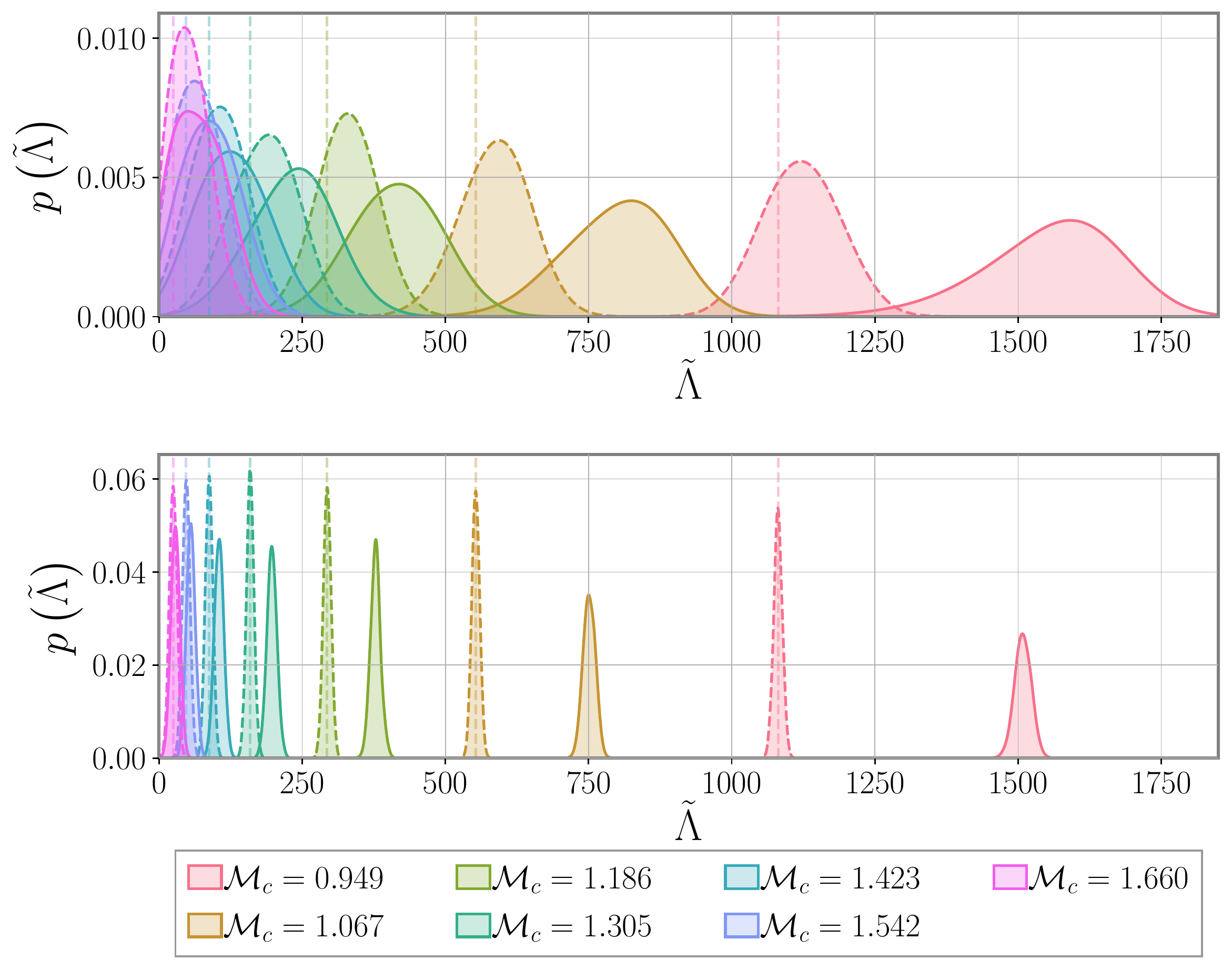}
\caption{One-dimensional posterior probability distributions of $\tilde{\Lambda}$ for BNSs with mass ratio $q=0.855$ and varying source-frame chirp mass $\mathcal{M}_c$ with the APR4 EOS as measured in A+ (top) and ET (bottom). For lighter binaries, which have larger tidal deformabilities, we see significant biases between adiabatic (solid) and dynamical tidal (dashed) posteriors. Vertical dashed lines indicate the injected values. As $\mathcal{M}_c$ increases, the effects of dynamical tides become negligible. }
\label{fig:bias1}
\end{figure}

The one-dimensional posterior probability distributions of $\tilde{\Lambda}$ with (dashed) and without (solid) the inclusion of dynamical tides in the recovery waveform model are shown in Fig.~\ref{fig:bias1} for the A+ network (top panel) and ET (bottom panel). As anticipated from the discussion above, dynamical tides play an increasingly negligible role for larger chirp masses (i.e. heavier NS and smaller $\tilde{\Lambda}$), but their neglect leads to large induced biases in the tidal measurements the lighter the NS. For the A+ network we find that the statistical uncertainties dominate for heavy BNS but for typical SNRs in ET, the tidal measurements of individual binaries are dominated by systematic errors. 

Tidal effects are larger (smaller) for stiffer (softer) EOS. The results for the soft APR4 EOS are the most conservative; for a medium-soft (SLy230A~\cite{Reinhard:1995rf}), and a medium-stiff (MPA1~\cite{Muther:1987mp}) EOS the impact of neglecting dynamical tides is even more prominent leading to even larger biases in the inferred $\tilde{\Lambda}$, especially for lighter NS. In all cases, tides are overestimated in order to compensate for the lack of dynamical tides, while simultaneously preferring more equal masses, see supplement for details.

\subsection{BNS population}
\label{sec:pop}
\begin{figure}[t]
    \centering
    \includegraphics[width=\columnwidth]{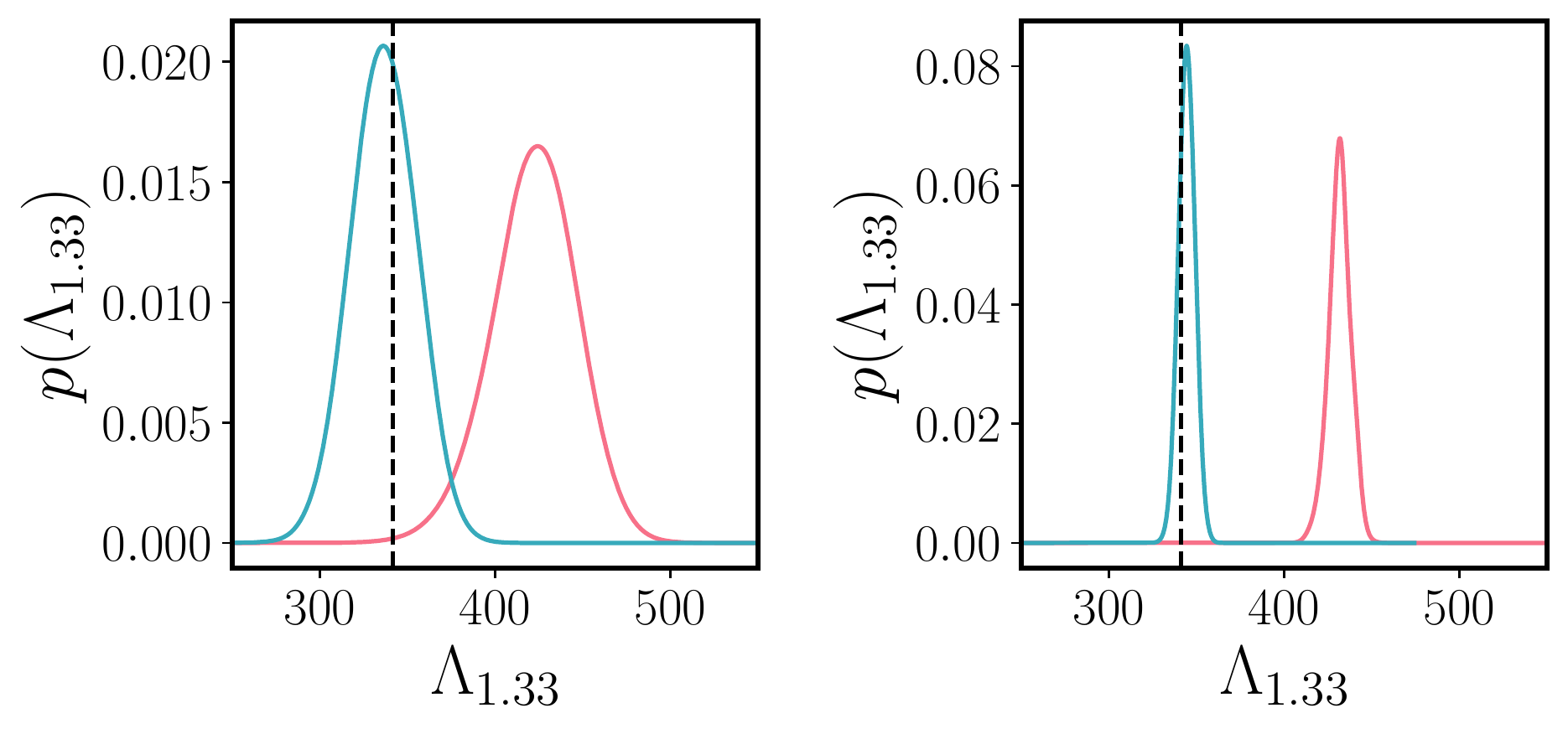}
    \caption{Stacked likelihood of the tidal deformability for $12$ BNS events in A+ (left) and ET (right) with adiabatic and dynamical tides (teal) and adiabatic tides only (red). The vertical dashed lines indicate the true value $\Lambda_{1.33}^{\rm true}=341$ assuming APR4 as the common EOS of our population.}
    \label{fig:likelihood}
\end{figure}
Whilst the bias in the tidal deformability can be less prominent on an event-by-event basis, it is manifestly evident at the population level and, strikingly, already observable in the A+ detector network. 

\begin{figure*}[t!]
\includegraphics[width=2\columnwidth]{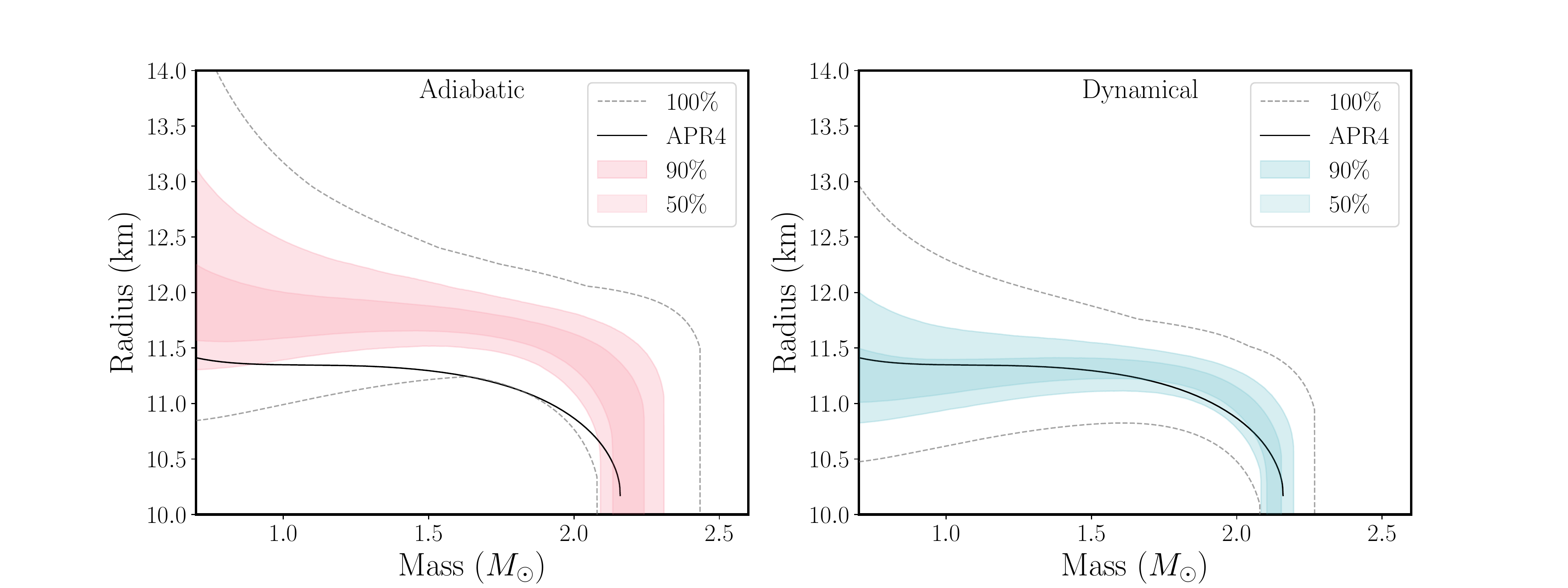}
\caption{Mass-radius relation for $12$ BNS observed in an A+ network. The population consists of binaries drawn from a Gaussian distribution consistent with the low-mass peak of galactic BNS, $m \sim \mathcal{N}(\mu_i,\sigma_i)$, where $\mu_i = 1.33M_{\odot}$. We use APR4 as the underlying equation of state, denoted by the black curve (black solid). Neglecting dynamical tidal effects induces biases $\sim \mathcal{O}(1\rm{km})$ in the inferred the mass-radius relation, excluding the true EOS at $90\%$ credible interval for almost all NS masses. 
}
\label{fig:mass_radius}
\end{figure*}

We now consider a prototypical BNS population with component masses drawn from a Gaussian distribution $m_ i \sim \mathcal{N} (\mu_i , \sigma_i)$, where $\mu_i = 1.33 M_{\odot}$ and $\sigma_i = 0.09 M_{\odot}$ \cite{Ozel:2016oaf}, consistent with the observed low-mass peak of binary neutron stars in the Milky Way~\cite{Tauris:2017omb}. Our knowledge of the true astrophysical population of BNS is still highly uncertain and current constraints possibly hint at a flatter distribution than that used here \cite{Landry:2021hvl}. Given the uncertainties in both GW observations and population synthesis models, a mass distribution following the Galactic population is a reasonable step towards understanding the impact of dynamical tides in a more realistic set of observations. We distribute the binaries randomly in the sky, uniformly in comoving volume between $40 \rm{Mpc}$ and $250 \rm{Mpc}$, and uniformly in inclination between $0^\circ$ and $30^\circ$, corresponding to binaries that are expected to be prime candidates for the detection of an associated electromagnetic (EM) counterpart~\cite{Rezzolla:2011da,Scolnic:2017sc,Cowperthwaite:2018gmx,Hosseinzadeh:2019ifm}. This also allows us to assume the detection of an EM counterpart and hence a known sky position for each binary, which in turn reduces the dimensionality of the concomitant Bayesian inference. 
Given the current median BNS merger rate~\cite{LIGOScientific:2020kqk}, for the A+ network we anticipate $\sim 30$ BNS detections with a distance $\leq 250$Mpc within a two-year observing window~\cite{Chen:2017wpg}, with 40\% being a potential target of opportunity for the Vera C. Rubin Observatory~\cite{Cowperthwaite:2018gmx}. Therefore, for our analyses we use $12$ BNS, broadly consistent with rate of joint GW-EM observations reported in \cite{Chen:2020zoq}. We consider both detector configurations and, as before, assume the soft APR4 EOS for all binaries. 

We first combine the information from the mass and tidal parameters of the individual BNS sources using the Taylor expansion method given by Eq.~\eqref{eq:stacked}. In Fig.~\ref{fig:likelihood}, we show the joint-likelihood for $\Lambda_{1.33}$ using $12$ BNS in the A+ detector network (left) and ET (right). We see that when dynamical tides are included in the recovery waveform model (teal), the true value $\Lambda_{1.33}^{\mathrm{true}} = 341$ is recovered within the 90\% credible interval in both detector networks, whereas neglecting dynamical tides (red) leads to a significant bias with median values of $423$ and $432$, respectively.

Due to the aforementioned limitations in EOS inference using the Taylor expansion of $\Lambda(m)$, we directly estimate the parameters of the common EOS following Eq.~\eqref{eq:eos_direct}.
Given a set of EOS parameters $\mathcal{E}$, we can solve the Tolman-Oppenheimer-Volkoff equations to determine the corresponding macroscopic properties such as the neutron star radius $R$ or the tidal deformability $\Lambda$. This allows us to map the mass and tidal parameters measured from a BNS observation to the parameterized EOS and vice versa. We again use $12$ BNS to infer the EOS parameters $\mathcal{E}$ with and without dynamical tides. The reconstructed mass-radius relations are shown in Fig.~\ref{fig:mass_radius}. As the population is drawn from a Gaussian centered on $1.33 M_{\odot}$, both the high-mass and low-mass behaviour of the EOS are poorly constrained. Critically, even in the A+ network we observe biases in the inferred radius of up to $\sim \mathcal{O}(1 \rm{km})$ if we fail to correctly account for dynamical tidal effects, translating into an incorrect preference for stiffer equations of state and the exclusion of the correct EOS at $90\%$ credible interval for nearly the entire NS mass range. 
A more comprehensive study on the role of dynamical tidal effects on the measurement of the EOS in future detector networks will be left to future work.

\section{Discussion}
Accurate modelling of higher-order tidal effects is a highly challenging endeavour and perforce results in approximations or simplifications being adopted. We have demonstrated that neglecting higher-order tidal effects, such as $f$-mode dynamical tides, can induce large biases in the measured tidal deformability and hence the inferred EOS. In particular, we have demonstrated how the systematic biases will pervade analyses of entire populations of observed BNS. Figure~\ref{fig:mass_radius} showcases how waveform models that omit dynamical tidal effects will potentially lead to $\mathcal{O}(1\rm{km})$ biases in the inferred radius of NS. Even more concerning is that these biases will already be relevant for A+ LIGO sensitivities, which are projected to be obtained in the fifth observing run ($\sim 2025$). These results suggest that the accurate modelling of higher-order tidal effects is a rather \textit{critical} problem that will lead to erroneous inferences about the EOS of NS if neglected. 

A caveat to our analysis is that we have assumed perfect knowledge of both the point-particle and adiabatic baselines. In reality, significant improvements will be required in the modelling of both effects as detector sensitivities continually improve~\cite{Samajdar:2018dcx, Gamba:2020wgg, Chatziioannou:2021tdi}. However, the point-particle baseline can be efficiently and effectively calibrated against state-of-the-art numerical relativity (NR) simulations \cite{Pratten:2020fqn, Ossokine:2020kjp}. Similarly, analytic modelling of adiabatic tidal effects has significantly improved in recent years \cite{Henry:2020ski, Nagar:2018zoe, Narikawa:2021pak} and the first waveform models that include dynamical tidal effects~\cite{Steinhoff:2016rfi,Hinderer:2016eia} have become available, but their usage in analysing data is severely limited due to their prohibitively large computational cost. 
Furthermore, our analysis only considers the impact of higher-order Newtonian tidal effects. Some of the current BNS waveform models include non-perturbative and non-linear tidal information to varying degree~\cite{Dietrich:2018uni, Dietrich:2019kaq}. However, due to the large computational cost of NR simulations, the model in \cite{Dietrich:2019kaq} was only calibrated to $\sim 4$ equal-mass NR simulations and is therefore limited in its validity and accuracy. An alternative model was presented in \cite{Kawaguchi:2018gvj} using a broader range of BNS simulations but restricting the model to a GW frequency of $f_{\rm max} = 1000 {\rm{Hz}}$. These NR calibrated models are equally susceptible to systematic errors but the clean separation into adiabatic and dynamical contributions becomes less transparent.
In summary, while some of the state-of-the-art waveform models already contain dynamical tides to some degree, their accurate analytical modelling as well as the calibration of waveform models against NR simulations incorporating the full mass ratio, spin and EOS dependence is urgently needed even, if very challenging, in order to accurately determine the NS EOS in near-future GW observations.

\section{Acknowledgments}
The authors thank Katerina Chatziioannou, Matt Nicholl, Lucy Thomas and Alberto Vecchio for useful discussions and comments.
G. P. and N. W. are supported by STFC, the School of Physics and Astronomy at the University of Birmingham and the Birmingham Institute for Gravitational Wave Astronomy.
P. Schmidt acknowledges support from the Dutch Research Council (NWO) Veni grant no. 680-47-460. 
Computations were performed using the University of Birmingham's BlueBEAR HPC service, which provides a High Performance Computing service to the University's research community, as well as resources provided by Supercomputing Wales, funded by STFC grants ST/I006285/1 and ST/V001167/1 supporting the UK Involvement in the Operation of Advanced LIGO.
This manuscript has the LIGO document number P2100307.

\bibliography{References}


\clearpage
\newpage

\setcounter{equation}{0}
\setcounter{figure}{0}
\setcounter{table}{0}
\setcounter{page}{1}

\title{Supplemental material: ``Impact of Dynamical Tides on the Reconstruction of the Neutron Star Equation of State''}
\maketitle

\onecolumngrid

Here we provide additional material to demonstrate the impact of dynamical tides on the GW phasing of inspiralling BNS and the measurement of the EOS. 

\section{GW Phasing}
\begin{figure}[h!]
    \centering
    \includegraphics[scale=0.5]{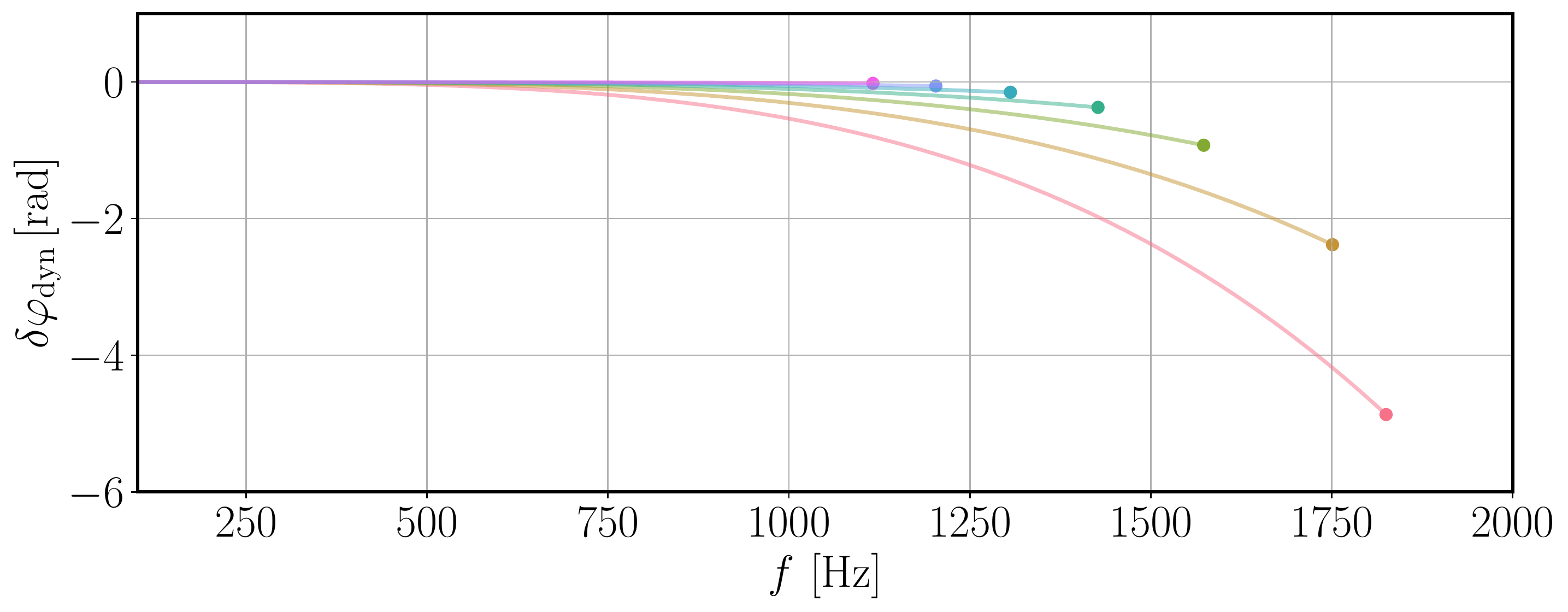}
    \caption{PN tidal phase associated with dynamical tides for the seven binaries of the systematic series with the soft APR4 EOS. As dynamical tides enhance the tidal interaction in the binary, the inspiral is accelerated relative to a binary with only adiabatic tides. The curves end at either the frequency of the innermost stable circular orbit or the estimated merger frequency~\cite{Dietrich:2018uni}.}
    \label{fig:phase}
\end{figure}

Despite the high PN order of dynamical tides~\cite{Schmidt:2019wrl} and their significant growth towards merger, their affect on the tidal GW phase can be up to a few radians even in the inspiral, depending on the masses of the neutron stars and the EOS. To illustrate the size of the effect, in Fig.~\ref{fig:phase} we show the PN tidal phase associated with dynamical tides, $\delta \varphi_{\rm dyn}$, for the systematic BNS series discussed in the main text. 

\section{Additional EOS}
\begin{figure*}[h!]
    \includegraphics[width=0.48\textwidth]{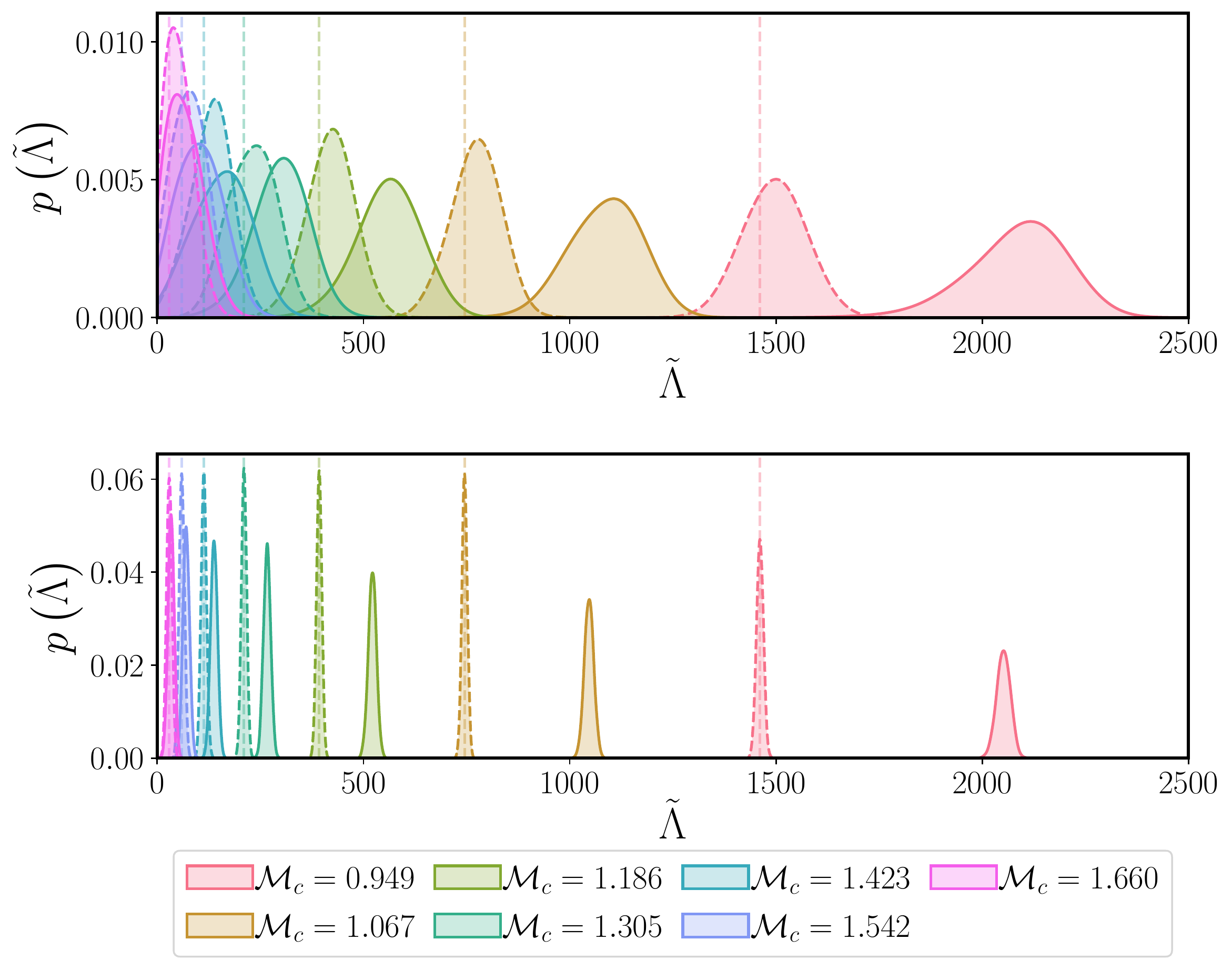}
    \includegraphics[width=0.48\textwidth]{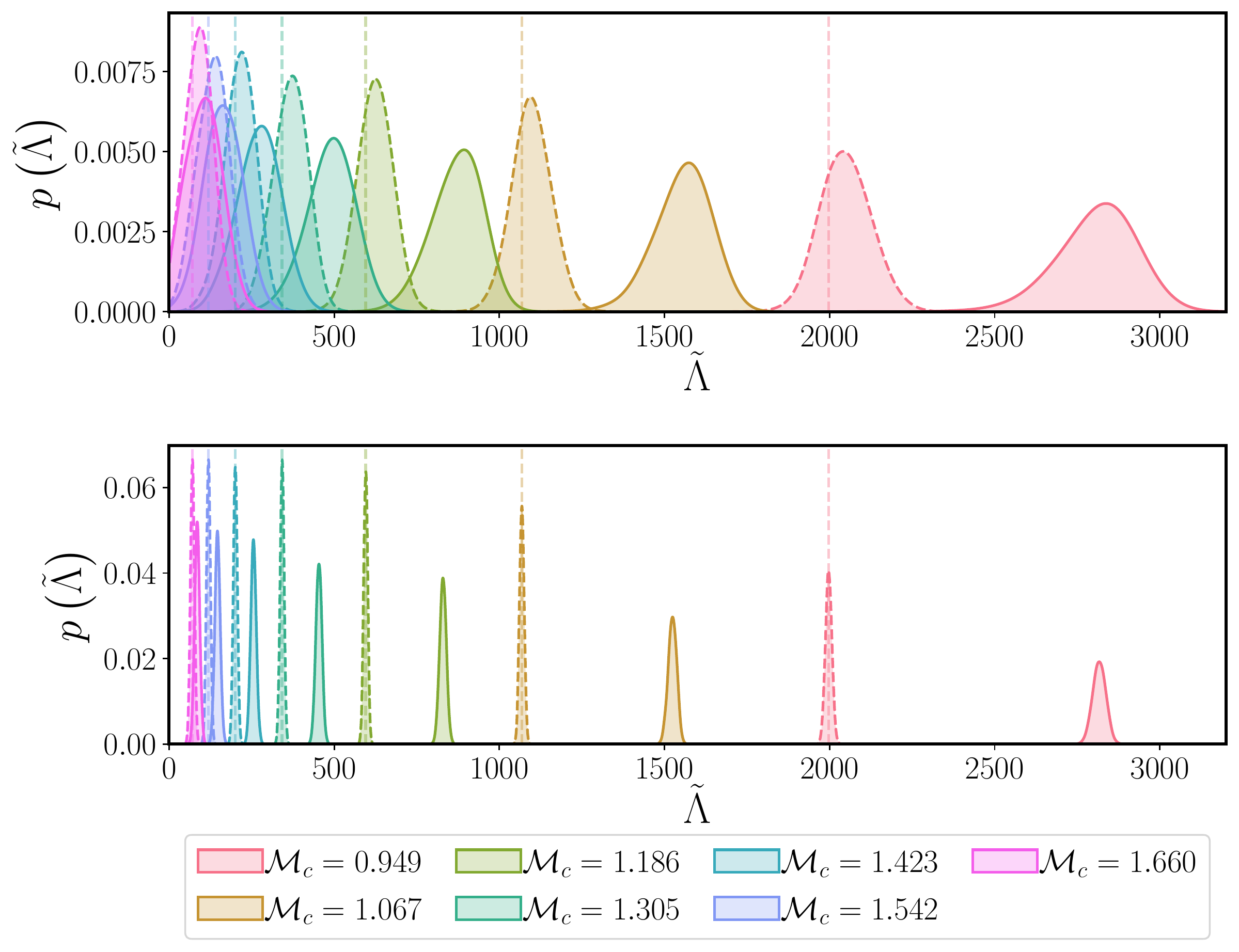}
    \caption{One-dimensional posterior probability distributions of $\tilde{\Lambda}$ for BNSs with mass ratio $q=0.855$ and varying source-frame chirp mass $\mathcal{M}_c$ with the SLy EOS (left) and the MPA1 EOS (right) as measured in A+ (top) and ET (bottom). For lighter binaries, which have larger tidal deformabilities, we see significant biases between adiabatic (solid) and dynamical tidal (dashed) posteriors. Vertical dashed lines indicate the injected values. As $\mathcal{M}_c$ increases, the effects of dynamical tides become negligible but the stiffer the EOS the larger the larger is the larger is observed bias.}
    \label{fig:MPA}
\end{figure*}

While the main text shows the results for the soft APR4 EOS, below we also show results for the medium-soft SLY230A and the stiff MPA1 EOS in Fig.~\ref{fig:MPA}. For a given BNS, the increase in the bias in the tidal deformability due to the neglect of dynamical tides is correlated with the stiffness of the EOS, i.e. the stiffer the EOS the larger the impact of dynamical tides or the neglect thereof due to the overall enhancement of tidal interactions. 

\section{Parameter Estimation}
\begin{figure}[h!]
    \centering
    \includegraphics[scale=0.5]{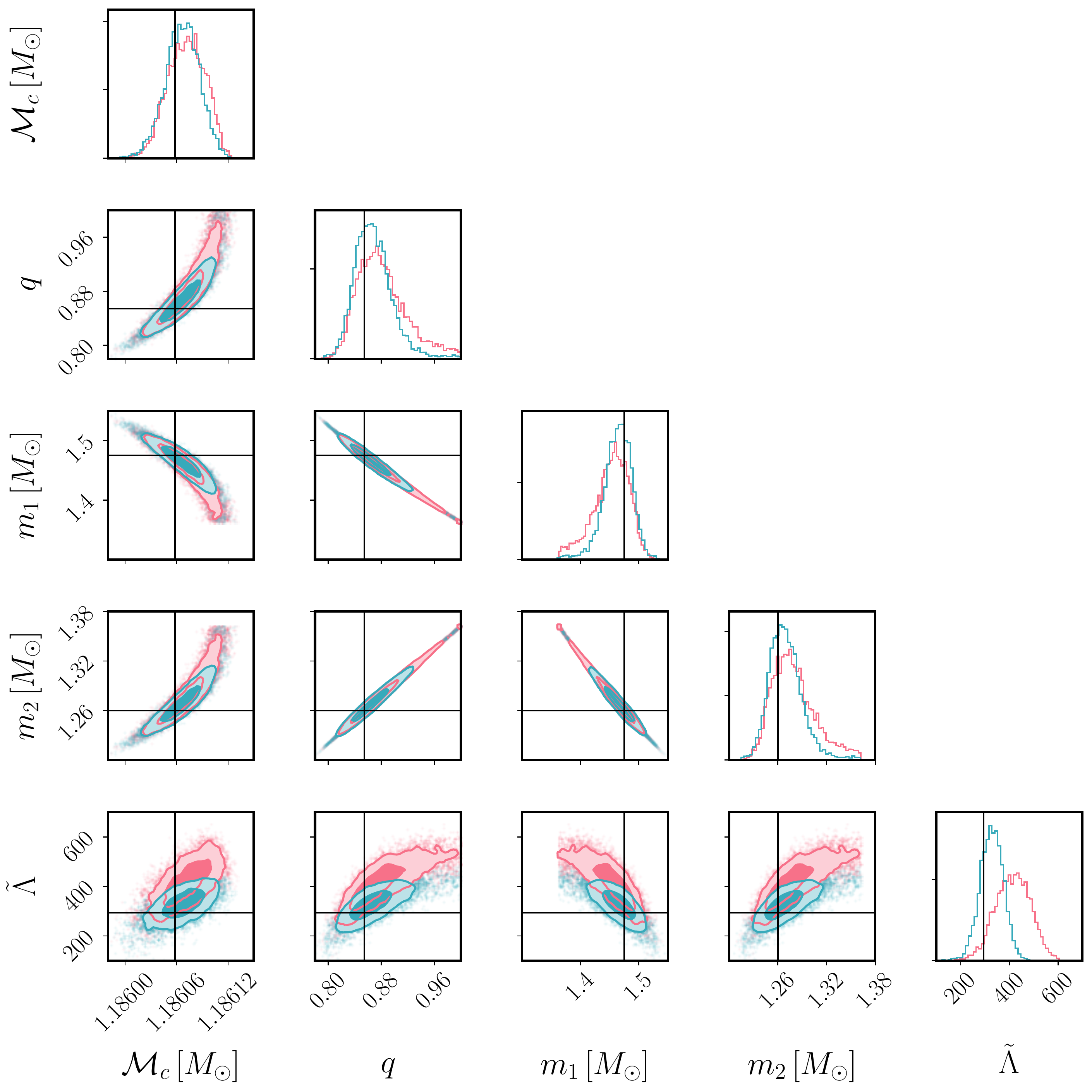}
    \caption{One- and two-dimensional posterior distributions for the mass parameters and the binary tidal deformability for the third binary of the systematic series with the soft APR4 EOS for the O5 sensitivity with (teal) and without (red) dynamical tides. We note that the sharp cut-offs in the component mass posteriors correspond to the equal mass limit.}
    \label{fig:corner}
\end{figure}

In Fig.~\ref{fig:corner} we show as an example the posteriors distributions obtained for the lightest binary of the systematic series for the O5 sensitivities with (teal) and without (red) dynamical tides. We see that the neglect of dynamical tides not only leads to an overestimation of the tidal deformability $\tilde{\Lambda}$ but also to biases in the mass parameters. Extrinsic parameters such as the inclination and polarisation are not affected. This is expected as the tidal terms in the GW phase are independent of those parameters. On the contrary, however, the adiabatic and dynamical tides both depend on the mass ratio and the component masses (see Eq.~(10) and Eq.~(2) of Refs.~\cite{Flanagan:1997fn} and~\cite{Schmidt:2019wrl}, respectively). We find that (i) the binary tidal deformability is overestimated, (ii) a more equal mass ratio is preferred and (ii) the component mass of the lighter NS is overestimated while the mass of the heavier NS is underestimated when dynamical tides are omitted. We observe this behaviour universally for all BNS considered here, which can qualitatively be understood as follows: Dynamical tides further accelerate the inspiral, changing the inspiral rate. Therefore, in their absence, larger tides are required to compensate for this difference. This is achieved by increasing the tidal deformability. Since we use universal relation to determine the $f$-mode frequency, $\Omega_{2A}$ decreases simultaneously. The leading-order dynamical tides contribution is $\propto (\Lambda_{2A} m_A^6)/(\eta \Omega_{2A}^2)$, where $\eta$ is the symmetric mass ratio. Lower component masses and a more equal mass ratio further enhance the tides. Since the chirp mass $\mathcal{M}_c$ is measured exquisitely well from the long inspiral, the remaining degree of freedom that can be adjusted to compensate for the missing dynamical tides is the mass ratio as can be seen in Fig.~\ref{fig:corner}.

\end{document}